# The SAT solving method as applied to cryptographic analysis of asymmetric ciphers

R.T. Faizullin[1], I.G. Khnykin[1], V.I.Dylkeyt[1]

[1] Omsk State Technical University, Russia, r.t.faizullin@mail.ru

**Abstract.** The one of the most interesting problem of discrete mathematics is the SAT (satisfiability) problem. Good way in SAT solver developing is to transform the SAT problem to the problem of continuous search of global minimums of the functional associated with the CNF. This article proves the special construction of the functional and offers to solve the system of non-linear algebraic equation that determines functional stationary points via modified method of consecutive approximation. The article describes parallel versions of the method. Also gives the schema of using the method to important problems of cryptographic analysis of asymmetric ciphers, including determining the concrete bits of multipliers (in binary form) in large factorization problems.

**Keywords:** CNF, SAT, resolution, minimization, cryptographic analysis, factorization.

## 1 Introduction

The subject of cryptographic analysis is of great significance, because strong ciphers are the foundation of modern telecommunication and financial systems. Progress in cryptographic analysis stimulates evolution in adjacent fields of mathematics such as algebra, theory of number, discrete mathematics.

Nowadays approaches to solve a problem of crypto strength of asymmetric cryptographic algorithms are either variations of number sieve algorithms [12] or Pollard's $\rho$ and $\lambda$ algorithms [11]. Periodical reports only confirm crypto strength of famous crypto algorithms. For instance, factorization of "large" numbers (~1000 bits) requires some months of supercomputer's machine time. And so, increasing key length solves the problem of crypto strength in principle.

The new approach to solve a problem of crypto strength is so called logical crypto analysis. Main point of logical crypto analysis is submission a crypto algorithm as a program for Turing's machine. Using plain and cipher texts in the program reduces the problem of crypto strength to the satisfiability problem (SAT). The part of satisfying set is the secret key of the crypto algorithm. For the first time, this idea was mentioned in [8]. Experience demonstrates that DP, DPL based algorithms with or without backtracking [10] are hard to use because of huge quantity of variables in associated CNF. And so, continuous methods should be used. In such methods the satisfying sets are associated with global minimums of the specific functional. For the first time, this idea was used in [3, 4]. There were attempts to associate the minimum search problem with physics models. In [5] we can see a chemical kinetics analogy, and in [6] there is a gravitation analogy. Note that continuous methods approach each variable in every step. Also, there is well-known that we can find minimum of the functional efficient if this minimum is singular and there are no more singular points. On the other hand, it is enough to find a set of bits that will congruent with the set of key bits with probability more than 0.5. And so, the searching of global minimum becomes a new and necessary test for crypto algorithms.

We can hope that numerous experiences in calculus and discrete mathematics will help to find close neighborhood of the key bit set.

## 2 Reducing SAT to minimization problem

Let's consider a CNF:
$$L(y) = \bigwedge_{i=1}^{M} c_i(y), \text{ where}$$
$$y \in B^N\{0,1\}, c_i(y) = \bigvee_{j \in \{1...N\}}^{M} l(y_j), \ l(y_j) = y_j \text{ or } \bar{y}_j$$

Associate $x$ with $y$, $(1-x)^2$ with $\bar{y}$. Here, $x \in R^N[0,1]$, $y \in B^N\{0,1\}$.

Let's reduce SAT to a global minimum searching problem:
$$\min_{x \in R^N[0,1]} F(x) = \sum_{i=1}^{M} C_i(x), \text{ where}$$

$$C_i(x) = \prod_{j=1}^{N} Q_{i,j}(x_j), \text{ where } Q_{i,j}(x_j) = \begin{cases} x_j^2, & \text{if } \bar{y}_j \in c_i(x) \\ (1-x_j)^2, & \text{if } y_j \in c_i(x) \\ 1, & \text{otherwise} \end{cases} \quad (1)$$

Here $\begin{cases} y_i \vee y_j \to x_i + x_j \\ y_i \wedge y_j \to x_i^2 x_j^2 \\ \bar{y}_i \to (1-x_i) \end{cases}$, where $\{y_i \in B, x_i \in R\}$

Note that $\min_{x \in R^N[0,1]} F(x) = 0$ equals to SAT.

Differentiating $F(x)$ by all $x_i$ we will have a non-linear system:
$$\sum_{\xi \in \Xi} z_j^2 z_k^2 ... \cdot x_i = \sum_{\xi \in \Lambda} z_j^2 z_k^2 ..., \ i = 1,2,..P, \text{ where}$$

$$z_i = \begin{cases} x_i, & \text{if } \bar{y}_i \in c_i(y) \\ (1-x_i), & \text{if } y_i \in c_i(y) \end{cases}$$

$$\Xi = \{\xi, i \in \xi : x_i \in c_i(x)\} \quad (2)$$

$$\Lambda = \{\xi, i \in \xi : \bar{x}_i \in c_i(x)\}$$

As shown in [1] the Newton's method is not efficient to find a decision of (2), because the decision belongs to kernel of operator $\frac{dF}{dx}(x)$. Because of this, so called, method of successive approximation with 'inertia' (SAI) (3) is offered:

$$\left[\sum_{p=0}^{K}\sum_{\xi \in \Xi} \alpha_k x_i(t-p)^2 x_j(t-k)^2\right] \cdot x_k(t+1) =$$

$$\sum_{\xi \in \Lambda} x_j^2(t) x_k^2(t) \underset{def}{\sim} A \cdot x_i(t+1) = B \quad (3)$$

$$\sum_{p=1}^{K} \alpha_p = 1, \ \alpha_p \in R[0,1]$$

Here iterations deal with real numbers but result vector is reduced to Boolean vector and this vector is used to examine the satisfiability.

Different modifications to improve the efficiency of SAI are shown below. The SAI algorithm with all modifications was called 'SAI mix'.

## 3 SAI mix

Initial CNF is transformed via resolution [7]. The resolution method transforms CNF to a CNF with fewer amounts of variables and disjuncts. The method computing complexity is $O(n \cdot \log(n))$.

The main procedure consists of iterations that combine SAI (3) and anti-gradient shifting. The right part of (2) is a gradient of *F(x)*, but decisions of (2) is stationary points of *F(x)* only. For instance, by random generating of CNF in such a way that the given set of bits be a decision of SAT, we get a CNF where amount of variables is nearly equal to amount of its negotiations. It means that *F(x)* have a 'kvazi-stationary' point $\{x_i = 0 \mid i = 1..N\}$ because of $A_i \sim 2B_i$. In practice the amount of variables is essentially not equal to amount of its negotiations and we can't guess 'kvazi-stationary' points. Each 'kvazi-stationary' point is corresponded to decision of indefinite systems that can be generated by excluding some equations from the initial system (2). The amount of such points increase exponentially and iteration procedure that search the point of minimum doesn't converge.

Basic iteration combines SAI (3) using Zeidel scheme and anti-gradient shifting: $x(t+1) = 2 \cdot x(t) - B/A$. [1]

While approaching the satisfying set, the rate of convergence can decrease. One of the reason is that the trajectory made by successive approximations trapped in areas of local minima. So called, 'changing trajectory method' lets escape the area of local minimum. The main idea of changing trajectory method is to form new 'better' approximation vector based on CNF and previous approximation that lets to continue searching [1].

## 4 Parallel version of SAI

Let's transform initial CNF into DNF. Divide the DNF into 2 independent parts (sub-formulas). Table 1 show that via SAI we can easily find the decision for each part. Each decision for sub-formulas $x_1 = \{x_{11}...x_{1N}\}$, $x_2 = \{x_{21}...x_{2N}\}$ is a point in $R^N$. Let's consider points in $R^N$:

$$x_l = \{x_{li} = \min(x_{1i}, x_{2i}) + \frac{|x_{1i} - x_{2i}|}{k} l \mid i = 1...N\}$$

We choose the point: $x_l : F(x_l) \underset{l}{\rightarrow} \min$ as a next approximation for SAI for initial CNF. Each sub-formula is processed at the same time.

This procedure help us to find 'the nearest' vector to the decision. There are only 2% of disjuncts left unsatisfied. And 2.5% of variables left 'undetermined', e.g. every satisfied disjunct remain satisfied independently of value of these variables.

## 5 SAI testing

There are some test benchmarks were chosen to test SAI: SAT 2005 benchmark [13], SATLib benchmark [14], random generated CNFs. Results are shown in table 1 and table 2.

**Table 1.** Results of parallel version of SAI (SATLib benchmark).

| Benchmark | Amount of variables (N) | Amount of disjuncts (M) | Amount of solved tests, % | Amount of iterations need to handle: | |
|---|---|---|---|---|---|
| | | | | Part of DNF | Whole CNF |
| RTI | 100 | 429 | 100 | 10 | 14 |
| BMS | 100 | <429 | 100 | 7 | 14 |
| sat05-1663 | 2000 | 8400 | 99 | 20 | 200 |
| sat05-1676 | 4000 | 16800 | 99 | 20 | 200 |
| sat05-1656 | 12000 | 50400 | 99 | 20 | 200 |
| UF20-91 | 20 | 91 | 100 | 10 | 14 |
| UF250-1065 | 250 | 1065 | 100 | 20 | 21 |

**Table 2.** Results of SAI (SATLib benchmark)

| Benchmark | Amount of variables (N) | Amount of disjuncts (M) | Amount of tests in benchmark | Amount of solved tests, % | Maximum amount of iterations |
|---|---|---|---|---|---|
| Backbone-minimal Sub-instances, 3-SAT ||||||
| RTI | 100 | 429 | 500 | 98,6 | 19988 |
| BMS | 100 | <429 | 500 | 79,8 | 29831 |
| Controlled Backbone Size Instances, 3-SAT ||||||
| CBS_b10 | 100 | 403 | 1000 | 100 | 38972 |
| CBS_b10 | 100 | 449 | 1000 | 100 | 38880 |
| CBS_b90 | 100 | 449 | 1000 | 98 | 29738 |
| Uniform Random 3-SAT (UF) ||||||
| UF20-91 | 20 | 91 | 1000 | 100 | 448 |
| UF250-1065 | 250 | 1065 | 100 | 98 | 9731 |
| Graph coloring problem ||||||
| FLAT30-60 | 90 | 300 | 100 | 100 | 4317 |

## 6 Using SAI in cryptographic analysis of asymmetric ciphers, summary

CNF associated with factorization problem, discrete logarithm problem, and discrete logarithm on elliptic curve problem are discussed in [2]. Estimated amount of disjuncts in CNF associated with factorization problem is nearly $CN^2$, $C \approx 10$, where $N$ is amount of bits. Estimated amount of disjuncts in CNF associated with discrete logarithm problem, and discrete logarithm on elliptic curve problem are $C_1 N^3, C_2 N^3$, $C_1 \approx 100, C_2 \approx 1000$ respectively. Resolution can twice reduce the amount of disjuncts for factorization problem and in 10 times for discrete logarithm problem, and discrete logarithm on elliptic curve problem.

The following results show how close SAI approximation vectors to the set (vector) of key bits of factorization problem are. We chose 30 test CNFs, associated with factorization of 1024, 2048, 3072 – bit numbers. Then approximation was compared with the vector of key bits at every iteration. 30 starts of SAI with random initial approximation were performed. The results are shown in fig. 1.

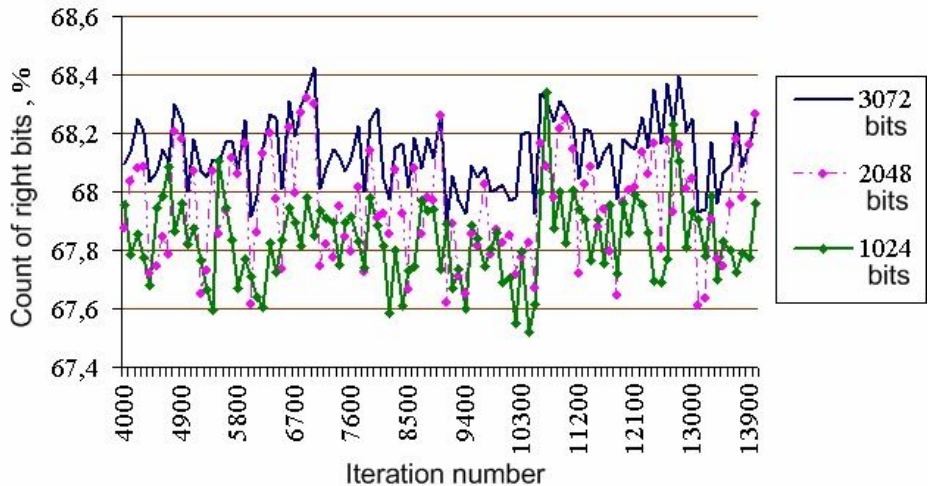

**Fig. 1.** Count of right bits of approximation vector (%) depending on iteration.

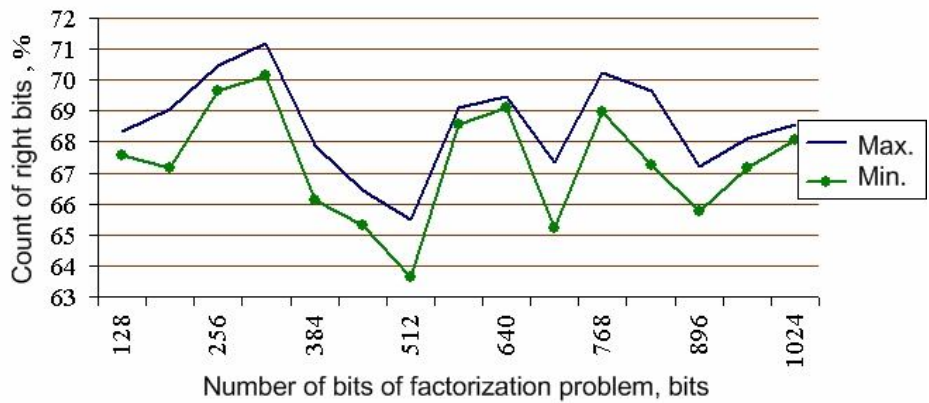

**Fig. 2.** Maximum and average count of right bits of approximation vector (%) depending on key size.

Fig. 1 show that nearly 68% of bits of approximation vector are always right in average. Maximum and minimum count of bits of approximation vector is always right in average too and estimated as 68.3%, 67.7% respectively. Note that we always get this result after 500-1000 iterations.

Fig. 2 show maximum and average count of right bits of approximation vector depending on key size. Note that these bits are significant, e.g. if we put values of variables into CNF, the SAT problem become easy to solve via SAI.

Fig. 3 show the matrix of long multiplication of binary numbers p and q after applying these tests.

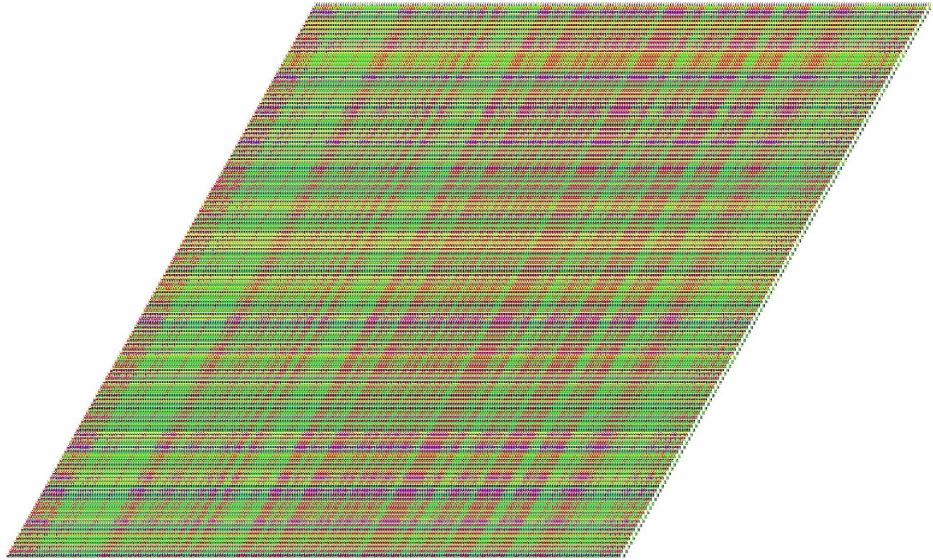

**Fig. 3.** Matrix of long multiplication of binary numbers p and q. Green - true bits, Blue, yellow - improved bits, Red - false bits.

Confidence intervals calculated via student's t-test are less then 0.18. Therefore, we can say that SAI have stable behavior in such kind of test.

Therefore, SAI form a set of bits that is close to set of key bits (on 68%). To determine which bits are right with high probability we offer a system of tests.

One of these tests consists in follow. Let us consider the matrix of long multiplication of binary numbers $p$ and $q$. Every line of this matrix is either $\bar{0}$ or $p$. Every column of this matrix is either $\bar{0}$ or $q$. Resolving this matrix by approximation vectors and comparing it with $p$, $q$ or $\bar{0}$ we can determine the values of unknown variables with certain probability. Reiterate this procedure with different approximations we can construct voting method for determining specific bits.

Table 3 show probability of determining right bits via test circumscribed above. The test was performed for 31 independent CNFs associated with 512 – bit factorization problem. Therefore, with probability of 0.8 bits in position (probability) 1(100%), 255(100%), 256(96%), 512(83%), 13(80%), 46(77%), 73(77%), 86(74%), 142(74%), 217(74%) for each factor are determined.

**Table 3.** Determining concrete key bits in 512 – bit factorization problem. Test 1 – Clusterization of long multiplication matrix (Count of tests is 31)

| Count of tests in which variables was determined | Count of tests in which variables was determined, % | Count of determined bits | Count of determined bits in % of 512 (count of all bits) |
|---|---|---|---|
| 31 | 100 | 2 | 0,4 |
| 30 | 96,77 | 1 | 0,2 |
| 26 | 83,87 | 1 | 0,2 |
| 25 | 80,65 | 1 | 0,2 |
| 24 | 77,42 | 2 | 0,4 |
| 23 | 74,19 | 8 | 1,6 |
| 22 | 70,97 | 9 | 1,8 |
| 21 | 67,74 | 21 | 4,1 |
| 20 | 64,52 | 50 | 9,8 |
| 19 | 61,29 | 66 | 12,9 |
| Sum | | 161 | 31,45 |

Another test consists in comparing the value *F(x)* (1) after resolving CNF by corresponding variable. Table 4 show that usually $F(x^R) > F(x^W)$ where $x^R$ is approximation vector with fixed right value of one component $x_i$, where $i = 1(100\%)$, 255(100%), 256(96%), 512(83%), 13(80%), 46(77%), 73(77%), 86(74%), 142(74%), 217(74%) and $x^W$ is approximation vector with fixed wrong value. Mentioned percents are taken from table 4.

**Table 4.** Determining concrete key bits in 512 – bit factorization problem). Test 2 – Comparing the values of associated with CNF functional (Count of tests is 31)

| Test bit | Value of F(x) with fixed: | | Functional values difference |
|---|---|---|---|
| | right value of test bit | wrong value of test bit | |
| 13 | 261,2 | 263,7 | - 2,5 |
| 46 | 260,8 | 263,5 | - 2,7 |
| 73 | 263,0 | 265,0 | - 2,0 |
| 86 | 254,5 | 256,7 | - 2,2 |
| 101 | 255,0 | 257,3 | - 2,3 |
| 142 | 263,2 | 259,8 | + 3,4 |
| 217 | 263,7 | 266,9 | - 3,2 |

The same result was got with more than one fixed variable. Therefore, results can cast doubt on RSA algorithm because using circumscribed tests together with parallel version of SAI we can determine bits in position 1, 255, 256, 512, 13, 46, 73, 86, 142, 217 with very high probability.

Table 5 show probability of determining right bits via test 2 for discrete logarithm problem. The test was performed for 100 independent CNFs associated with , 88 – bit problem.

**Table 5.** Determining concrete key bits in discrete logarythm problem). Test 2 – Comparing the values of associated with CNF functional (Count of tests is 100)

| Number of tests with true determined bits (see right columns) | Number of true determined bits | | |
|---|---|---|---|
| | 48 bits | 64 bits | 88 bits |
| 82 | 0 | 1 | 0 |
| 80 | 0 | 0 | 0 |
| 78 | 0 | 0 | 1 |
| 76 | 0 | 0 | 0 |
| 74 | 1 | 0 | 0 |
| 72 | 0 | 0 | 1 |
| 70 | 0 | 1 | 0 |
| 68 | 0 | 0 | 1 |
| 66 | 4 | 0 | 2 |
| 64 | 0 | 0 | 1 |
| 62 | 2 | 1 | 4 |
| 60 | 1 | 3 | 4 |
| 58 | 3 | 4 | 5 |
| 56 | 5 | 4 | 12 |
| 54 | 3 | 4 | 9 |
| 52 | 4 | 3 | 10 |
| Sum | 23 | 21 | 50 |

The next aim of this work is to develop more tests and voting methods for determining specific bits.